\documentclass[conference]{IEEEtran}
\IEEEoverridecommandlockouts

\usepackage{cite}
\usepackage{amsmath,amssymb,amsfonts}
\usepackage{acronym}
\usepackage{subfigure} 
\usepackage{multirow}
\usepackage{booktabs}
\usepackage{algorithmic}
\usepackage{graphicx}
\usepackage{textcomp}
\usepackage{xcolor}

\def\BibTeX{{\rm B\kern-.05em{\sc i\kern-.025em b}\kern-.08em
    T\kern-.1667em\lower.7ex\hbox{E}\kern-.125emX}}

\input{AcronymList}
\begin{document}

\title{On the Performance of Handover Mechanisms for Non-Terrestrial Networks\\ 
\thanks{This work was supported by the Scientific and Technological Research Council of Turkey (T\"{U}B\.{I}TAK) under Grant 5200107, with the cooperation of Turkcell Technology and İstanbul Medipol University.
\par This work has been submitted to the IEEE for possible publication. Copyright may be transferred without notice, after which this version may no longer be accessible.}
}

\author{
\IEEEauthorblockN{Yusuf~Islam~Demir\IEEEauthorrefmark{1}\IEEEauthorrefmark{2},~Muhammad~Sohaib~J.~Solaija\IEEEauthorrefmark{1}, and H\"{u}seyin Arslan\IEEEauthorrefmark{1}\IEEEauthorrefmark{3}}\\
\IEEEauthorblockA{\IEEEauthorrefmark{1}Department of Electrical and Electronics Engineering, Istanbul Medipol University, Istanbul, 34810 Turkey\\
}
\IEEEauthorblockA{\IEEEauthorrefmark{2}Department of Electrical and Electronics Engineering, Istanbul University-Cerrahpa\c{s}a, Istanbul, 34320 Turkey\\
}
\IEEEauthorblockA{\IEEEauthorrefmark{3}Department of Electrical Engineering, University of South Florida, Tampa, FL 33620 USA\\
Email: yusuf.demir@medipol.edu.tr, solaija@ieee.org, huseyinarslan@medipol.edu.tr}
}

\maketitle

\begin{abstract}
Next-generation wireless networks require massive connectivity and ubiquitous coverage, for which non-terrestrial networks (NTNs) are a promising enabler. However, NTNs, especially non-geostationary satellites bring about challenges such as increased handovers (HOs) due to the moving coverage area of the satellite on the ground. Accordingly, in this work, we compare the conventional measurement-based HO triggering mechanism with other alternatives such as distance, elevation angle, and timer-based methods in terms of the numbers of HOs, ping-pong HOs, and radio link failures. The system-level simulations, carried out in accordance with the 3GPP model, show that the measurement-based approach can outperform the other alternatives provided that appropriate values of hysteresis/offset margins and time-to-trigger parameters are used. Moreover, future directions regarding this work are also provided at the end. 
\end{abstract}


\section{Introduction}
\label{Sec:Intro}
\par \Ac{5G} wireless networks have shifted the focus from high data rates to increased coverage, massive connectivity, high reliability, low latency, high mobility, and better power efficiency. All these requirements are expected to be escalated even further with the introduction of \ac{6G} networks \cite{zhang20196g, dang2020should}. As such, the deployment of \acp{NTN} is becoming an increasingly popular approach to address the connectivity outage as the users look for reliable and ubiquitous service irrespective of the location \cite{giordani2020non}. However, despite its advantages related to speedy and flexible deployment \ac{NTN} brings about its own unique challenges such as propagation channel and delay, frequency/bandwidth plan, link budgeting, mobility of the satellite, and \acp{UE} \cite{3GPP_38_811}. 

\par
\begin{figure}[t!]
   \centering
    \includegraphics[width=0.95\linewidth]{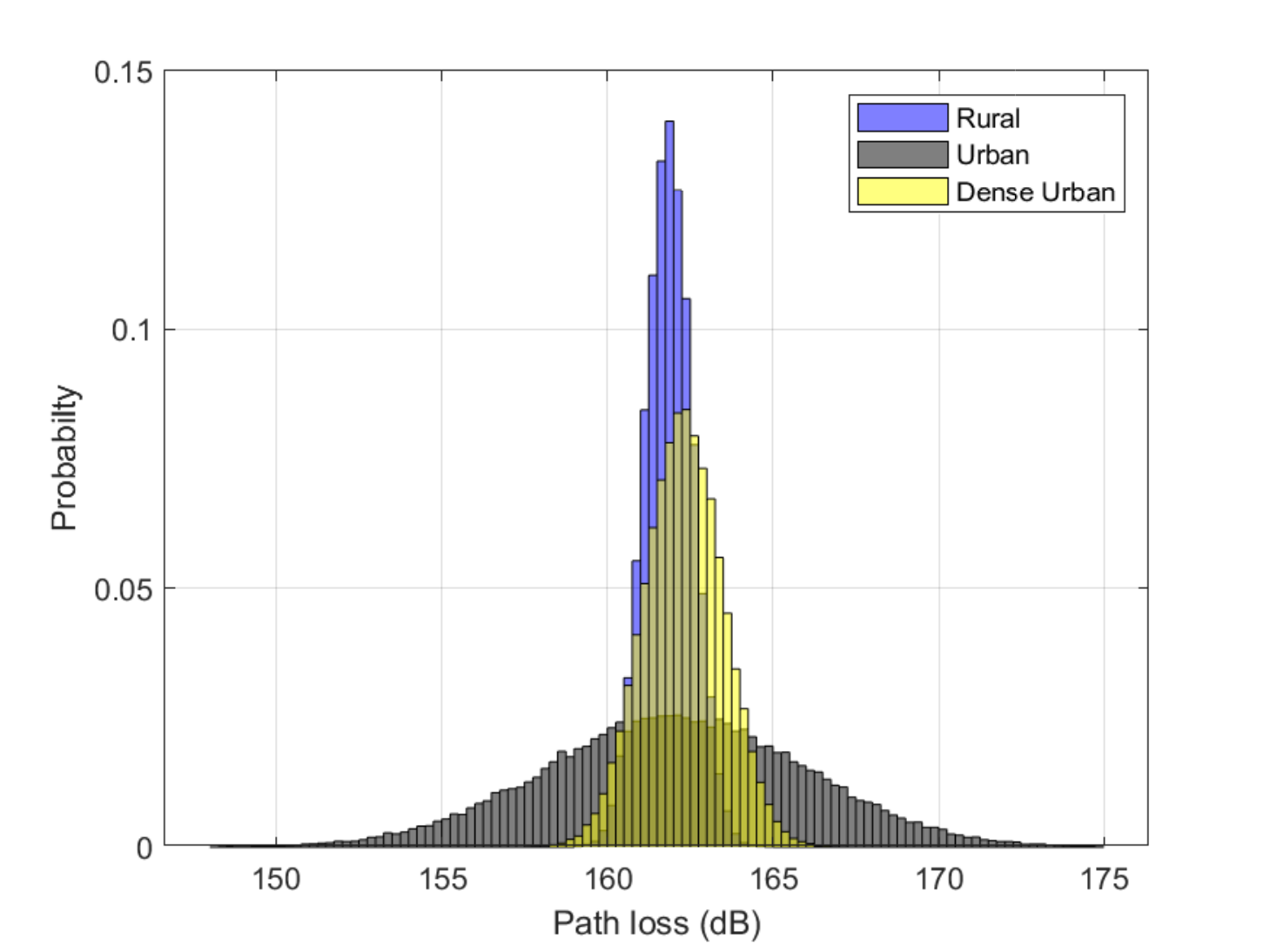}
    \caption{Comparison of the path loss for different environments for cell diameter of $50$km and \ac{NTN} altitude of $600$km.}
    \label{fig:RSS_variation}
\end{figure}

 Mobility is challenging even for conventional terrestrial networks because it leads to issues such as Doppler spreading, time selectivity of the channel, and increased \acp{HO} \cite{wu2016survey}. These problems are further compounded in \ac{LEO} satellites since their footprint can vary with speeds up to $7.56$ km/s \cite{3GPP_38_821} which is significantly faster than terrestrial user mobility (for comparison, high-speed trains move with speeds around  $0.14$ km/s). Even with the larger footprints of \acp{NTN}, their speed potentially causes an increased number of \acp{HO}. Moreover, the propagation distance from the satellite to ground can be much higher compared to coverage footprint, which would lead to reduced \ac{RSS} variation in these networks compared to terrestrial networks \cite{3GPP_38_821}. This predicament is illustrated for different environments in Fig. \ref{fig:RSS_variation}. While urban environment still exhibits an \ac{RSS} range of around $20$ dB, this variation is limited to $4-5$ dB in rural and dense urban environments.

\begin{figure*}[t!]
    \centering
    \includegraphics[width=0.95\linewidth]{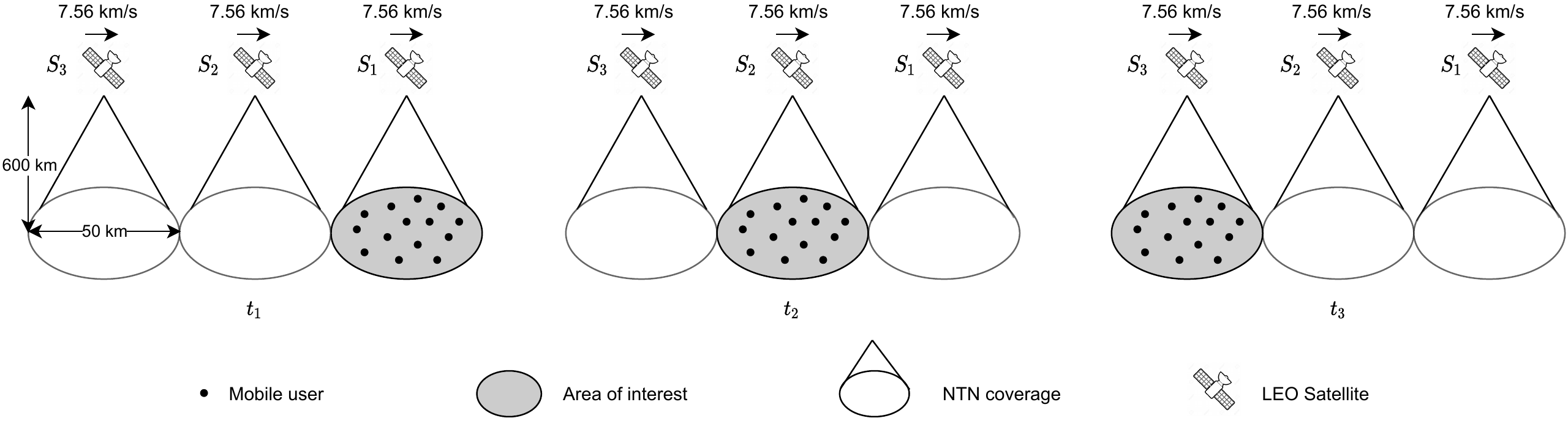}
    \caption{Illustration of the considered scenario.}
    \label{fig:NTN_scenario}
\end{figure*}

\par In conventional (terrestrial) networks, the users are handed over from one cell or eNodeB to another based on measured \ac{RSS}, \ac{RSRP}, or \ac{RSRQ}. Specifically, in the case of A3 events, the \ac{HO} is triggered when the received signal of the neighboring \ac{eNB} is a certain threshold (hysteresis) stronger than the serving one for a specific amount of time \cite{3GPP_38_331}. However, with the reduced \ac{RSS} variation in \acp{NTN} the efficacy of the measurement-based \ac{HO} mechanisms could be compromised. 

\par Accordingly, some recent works have looked at different aspects of this problem. The general trends related to \ac{LEO} satellites \ac{HO} mechanisms are provided in \cite{park2021trends}, with the authors highlighting the need for content and device-specific approaches in future networks. The performance of conventional \ac{HO} mechanism in \ac{LEO}-based \acp{NTN} is compared against two terrestrial scenarios, i.e., urban macro and high-speed train in \cite{juan20205g}, with the results showing that the performance is significantly degraded in terms of outage, \acp{HO} and \acp{RLF}. The achievable data rates/capacity for \ac{mmWave} band operation in \acp{NTN} is studied in \cite{giordani2020satellite}. In \cite{baik2021analysis}, authors suggest and analyze the use of fixed-beam \ac{LEO} satellites to mitigate the excessive \acp{HO} in the case of moving beams. Recent releases of \ac{3GPP} specifications have also considered \ac{CHO} as potential improvement of the conventional \ac{HO} methods, and \cite{martikainen2018basics} studies the effect of different values of parameters such as add, remove, replace, and execution offset and its impact on \acp{HO} and \acp{RLF}.  

\par However, none of the works mentioned above have evaluated any alternatives to the conventional \ac{HO} mechanisms so far. Accordingly, in this work, we first provide an implementation of the alternative handover mechanisms suggested in \cite{3GPP_38_821}. Moreover, we carry out a simulation-based comparative analysis of these mechanisms, i.e., distance/position, elevation angle, and timer-based methods. The performance of these methods is evaluated in terms of the number of \acp{HO}, \acp{PPHO}, and \acp{RLF} and benchmarked against the conventional measurement-based approach. 

\par The rest of this paper is organized as follows. Section \ref{Sec:SystemModel} describes the system model used in this study, the alternative \ac{HO} mechanisms are discussed in Section \ref{sec:altHO}, simulation results are provided in Section \ref{sec:Results} while conclusions and future directions are provided in Section \ref{sec:Conclusion}.

\section{System Model and Assumptions}
\label{Sec:SystemModel}
\par In this work we assume a single circular cell on the ground with outdoor users randomly distributed inside the cell, while three \ac{LEO} satellites move overhead, as shown in Fig. \ref{fig:NTN_scenario}. The simulation is assumed to start at time $t_1$ where satellite $S_1$ is right above the cell center. The \ac{LEO} high-speed motion is assumed to continue linearly at a speed of $7.56$ km/s till satellite $S_3$ is on top of the cell center. Since our primary goal is to evaluate the performance of the different \ac{HO} mechanisms in the challenging circumstances provided by \acp{NTN}, we have chosen the worst case, i.e., the lowest altitude and smallest cell diameter, which are $600$ and $50$ km. The low altitude leads to a fast-moving coverage footprint while the small cell size leads to reduced \ac{RSS} variation. In the remainder of this section, we will first describe the path loss model and its associated parameters before providing the details of the user mobility pattern used in our analysis.

\subsection{Path Loss Model}
\label{sec:pathloss}
The path loss model for \acp{NTN} was developed by \ac{3GPP} in Rel-15 \cite{3GPP_38_811}, and is reproduced below
\begin{equation}
\label{eq:PL_1}
    PL = PL_b + PL_s + PL_g + PL_e, 
\end{equation}
where basic path loss, ionospheric/tropospheric scintillation, attenuation due to atmospheric gasses, and building entry losses (all in dB) are represented by $PL_b$, $PL_s$, $PL_g$ and $PL_e$, respectively. The basic path loss is given by
\begin{equation}
\label{eq:PL_b}
    PL_b =  SF + CL + FSPL,
\end{equation}
where $SF\sim\mathcal{N}(0,\sigma_{SF}^2)$ represents the shadow fading, $CL$ represents the clutter loss and $FSPL$ represents the free-space path loss. The values of $\sigma_{SF}$ and $CL$ are given in Table \ref{tab:LoS_SF_CL}. The free-space path loss is given by
\begin{equation}
\label{eq:FSPL}
    FSPL = 32.45 + 20\log_{10}(f_c) + 20\log_{10}(d),
\end{equation}
where $f_c$ represents the carrier frequency in GHz and $d$ is the 3-D distance (in meters) between the \ac{UE} and satellite calculated by
\begin{equation}
\label{eq:dist}
    d = \sqrt{R_{E}^{2}\sin^2\alpha + h_0^2 +2h_0R_E} - R_E\sin\alpha,
\end{equation}
where $R_E$ is the radius of the Earth, $\alpha$ is the elevation angle and $h_0$ is the altitude of the satellite. 

\par For $fc<6$ GHz, tropospheric scintillation is considered to be negligible while ionospheric scintillation is equivalent to $P_{fluc}/\sqrt{2}$ where $P_{fluc}$ is a function of the $S_4$ scintillation parameter \cite{ITU_531_14}. The absorption due to atmospheric gases, $PL_g$, is assumed to be negligible for $fc<6$ GHz, while the building penetration loss, $PL_e$, is also ignored since all the users are assumed to be outdoors \cite{3GPP_38_811}. 

\par The overall path loss then becomes a weighted sum of the \ac{LoS} and \ac{NLoS} components and is given by \cite{al2014optimal}
\begin{equation}
\label{eq:PL_tot}
    PL_{tot} = Pr_{LoS}(\alpha)*PL_{LoS} + (1 - Pr_{LoS}(\alpha))*PL_{NLoS}, 
\end{equation}
where $Pr_{LoS}(\alpha)$ gives the \ac{LoS} probability for elevation angle $\alpha$, $1 - Pr_{LoS}(\alpha)$ is the \ac{NLoS} probability, and $PL_{LoS}$ and $PL_{NLoS}$ represent the path losses for the \ac{LoS} and \ac{NLoS} components, respectively.

\subsection{User Mobility Model}
\par In this work, both static and mobile users are considered. For mobile users, memory-based smooth random mobility model is used. As the name implies, in memory-based models a user's current state (in terms of speed and/or direction) is a function of its past state(s) which is more realistic compared to memory-less models such as random walk \cite{tabassum2019survey}. The model provided in \cite{mobility_user} considers user speeds in the interval $[0, v_{max}]$, where certain speeds have higher probability while uniform distribution is used in the remaining interval. The change in direction is done after a time interval which is itself exponentially distributed. For a more detailed explanation of the model, the readers are referred to \cite{mobility_user}. However, it should be noted that to provide a fair comparison with the static user case the user mobility had to be limited to the considered cell region. This was ensured by modifying the aforementioned model to change the user direction by a shift of $\pi$ radians if it reached the cell edge. 

\begin{table}[]
\centering
\caption{Path loss parameters for \acp{NTN} in dense urban scenario \cite{3GPP_38_811}}
\label{tab:LoS_SF_CL}
\begin{tabular}{|c|c|cc|c|}
\hline
{\color[HTML]{333333} }                                    &        & \multicolumn{2}{c|}{\textbf{Shadow Fading ($\sigma_{SF}$)}} &      \\ \cline{3-4}
\multirow{-2}{*}{{\color[HTML]{333333} \textbf{\begin{tabular}[c]{@{}c@{}}Elevation\\ Angle ($\alpha$)\end{tabular}}}} &
  \multirow{-2}{*}{\textbf{\begin{tabular}[c]{@{}c@{}}LoS\\ Probability\end{tabular}}} &
  \multicolumn{1}{c|}{\textbf{LoS}} &
  \textbf{NLoS} &
  \multirow{-2}{*}{\textbf{\begin{tabular}[c]{@{}c@{}}Clutter \\ Loss (CL)\end{tabular}}} \\ \hline
{\color[HTML]{333333} \textbf{10$^{\circ}$}} & 28.2\% & \multicolumn{1}{c|}{3.5}          & 15.5         & 34.3 \\ \hline
{\color[HTML]{333333} \textbf{20$^{\circ}$}} & 33.1\% & \multicolumn{1}{c|}{3.4}          & 13.9         & 30.9 \\ \hline
{\color[HTML]{333333} \textbf{30$^{\circ}$}} & 39.8\% & \multicolumn{1}{c|}{2.9}          & 12.4         & 29.0 \\ \hline
{\color[HTML]{333333} \textbf{40$^{\circ}$}} & 46.8\% & \multicolumn{1}{c|}{3.0}          & 11.7         & 27.7 \\ \hline
{\color[HTML]{333333} \textbf{50$^{\circ}$}} & 53.7\% & \multicolumn{1}{c|}{3.1}          & 10.6         & 26.8 \\ \hline
{\color[HTML]{333333} \textbf{60$^{\circ}$}} & 61.2\% & \multicolumn{1}{c|}{2.7}          & 10.5         & 26.2 \\ \hline
{\color[HTML]{333333} \textbf{70$^{\circ}$}} & 73.8\% & \multicolumn{1}{c|}{2.5}          & 10.1         & 25.8 \\ \hline
{\color[HTML]{333333} \textbf{80$^{\circ}$}} & 82.0\% & \multicolumn{1}{c|}{2.3}          & 9.2          & 25.5 \\ \hline
{\color[HTML]{333333} \textbf{90$^{\circ}$}} & 98.1\% & \multicolumn{1}{c|}{1.2}          & 9.2          & 25.5 \\ \hline
\end{tabular}
\end{table}

\section{Alternative Handover Realizations}
\label{sec:altHO}
In this section we describe the \ac{HO} triggering mechanisms, starting from the conventional \ac{RSS}-based methods before looking at other measurements such as distance, timer, and elevation angle which have been identified as potential \ac{HO} triggers in \ac{3GPP} discussions \cite{3GPP_38_821}. 

\subsection{Measurement-Based Handover}
In general, measurement-based \ac{HO} techniques rely on the signal strength of the received signal to trigger \acp{HO}. The specific measured quantity might change, for instance \ac{RSRP}, \ac{RSRQ} or \ac{RSSI} might be used \cite{ahmad2017handover}. While there are different variants of measurement-based \ac{HO} triggering events, we will limit ourselves to the A3 event, which is arguably the most popular. In addition to the measured signal strength, there are some additional parameters that are used in A3 measurements, i.e., hysteresis/offset margins and \ac{TTT}. Hysteresis and offset margins are used to bias the user cell association towards the serving cell by making it look stronger than its actual signal strength to avoid unnecessary \acp{HO}. In addition to this, \ac{TTT} parameter dictates how long the signal strength condition lasts to avoid \acp{PPHO}. Mathematically, the measurement-based \ac{HO} trigger can be written as

\begin{equation}
\label{eq:meas}
    Q_t > Q_s + Q_{hys} + Q_{off},
\end{equation}
where $Q$ can refer to any signal strength measurement such as \ac{RSRP}, \ac{RSRQ} or \ac{RSSI}. However, in this work we will focus on \ac{RSS}. The subscripts $t$ and $s$ refer to target and source cells, respectively, and $Q_{hys}$ and $Q_{off}$ are the respective hysteresis and offset margins. As mentioned earlier, the condition in \eqref{eq:meas} should be satisfied for a minimum of \ac{TTT} duration for the \ac{HO} to be triggered. The \ac{RSS} is calculated as follows:

\begin{equation}
\label{eq:RSS}
    Q = EIRP - PL_{tot},
\end{equation}
where \ac{EIRP} is a function of the transmit power and transmit antenna gain while $PL$ is the total path loss as described in Section \ref{sec:pathloss}

\subsection{Distance/Location-Based Handover}
In the distance-based approach, the initial association of the \acp{UE} is done based on the distance, i.e., the \ac{UE} associates itself with the nearest satellite. This is done with the assumption that the location of the \ac{UE} and \ac{NTN} satellite is exactly known and the distance can be calculated from that. Similar to the measurement-based approach, we use a hysteresis margin in the distance-based method while triggering the \ac{HO} condition. This can be illustrated as
\begin{equation}
\label{eq:distHO}
    d_t < d_s - d_{off},
\end{equation}
where $d_t$ and $d_s$ are the \ac{UE}'s distances from the target and source satellites, respectively, and $d_{off}$ is the offset parameter. 

\subsection{Elevation Angle-Based Handover}
Similar to the distance-based method, in elevation angle-based approach the \ac{UE} is associated with the satellite that has the largest elevation angle, $\alpha$. Here again we assume the availability of \ac{UE} and satellite's locations which can then be used to calculate $\alpha$. The former's location can be obtained using \ac{GPS} coordinates \cite{3GPP_38_821} while ephemeris data can be used for the latter \cite{liberg2020narrowband}. For \ac{HO} triggering, an offset parameter is used as follows:
\begin{equation}
\label{eq:angle}
    \alpha_t > \alpha_s + \alpha_{off},
\end{equation}
where $\alpha_t$ and $\alpha_s$ are the \ac{UE}'s elevation angles with respect to the target and source satellites, respectively, and $\alpha_{off}$ is the offset parameter.   

\subsection{Timer-Based Handover}
\par The timer-based \ac{HO} trigger is aimed at utilizing the knowledge of the satellite speed/trajectory to predict the time duration for which the satellite's footprint covers a certain cell on the ground. For instance, considering the satellite speed as $7.56$ km/s, a cell of $50$ km in diameter would be covered by a particular satellite for only $50/7.56=6.61$ seconds. As such, it is possible to initiate a timer once the user makes its initial association with a satellite and initiate the \ac{HO} after every $t_{off} \approx 6.61$ s.

\par As described in Section \ref{Sec:SystemModel}, we assume the satellite $S_1$ to be in line with the cell center at time $t_1$. As such, we cannot initiate the timer, $t_{off}$ with respect to first user cell association. To maintain fairness, we use the same methodology as distance-based \ac{HO} in the initial phase, i.e., before the \ac{HO} from $S_1$ to $S_2$. Once the \ac{UE} is handed over to $S_2$, timer $t_{off}$ is initiated and \ac{HO} to $S_3$ is initiated once the timer reaches its threshold. 

\begin{table}[t!]
\centering
\caption{Simulation assumptions and parameters}
\label{tab:assumptions}
\resizebox{0.65\columnwidth}{!}{
\begin{tabular}{|l|l|}
\hline
\textbf{Parameter}         & \textbf{Values}      \\ \hline
Environment                & Dense Urban          \\ \hline
NTN altitude ($h_0$)       & 600 km               \\ \hline
Cell diameter              & 50 km                \\ \hline
EIRP density               & 34 dBW/MHz           \\ \hline
Carrier frequency ($f_c$)  & 2 GHz (S-Band)       \\ \hline
Scintillation ($S_4$)      & 0.5                  \\ \hline
$P_{fluc}$ (dB)                 & 11                   \\ \hline
\begin{tabular}[c]{@{}l@{}}Radio link failure\\ (RLF) parameters\end{tabular} &
  \begin{tabular}[c]{@{}l@{}}$Q_{in}$ = $-6$ dB,\\ $Q_{out}$ = $-8$, dB\\ T310 timer = $500$ ms\end{tabular} \\ \hline
TTT (ms)                   & 20, 40, 60, 80, 100  \\ \hline
$Q_{hyst} +  Q_{off}$ (dB) & 1, 2, 3, 4           \\ \hline
$\alpha_{off}(^{\circ})$  & 1, 2, ..., 10        \\ \hline
$t_{off}$ (s)           & 6.4, 6.45, ..., 6.8 \\ \hline
$d_{off}$ (km)       & 1, 1.5, ..., 5     \\ \hline
$v_{max}$ (m/s)            & 10                   \\ \hline
\end{tabular}
}
\end{table}

\section{Simulation Results and Discussion}
\label{sec:Results}
For this work, we have considered downlink communication in S-band between \ac{LEO} \acp{NTN} and ground users in a dense urban scenario. In line with set-1 of the system-level simulation parameters provided in \cite{3GPP_38_821}, we assume \acp{LEO} at an altitude, $h_0$, of $600$ km and \ac{EIRP} density of $34$ dBW/MHz. Users are randomly located with a density of $\sim1$ UE/km$^2$ in the cell, where each user is allocated a single \ac{PRB} throughout the simulation. The cell size is chosen to be equal to the satellite beam diameter, i.e., $50$ km. The noise in the system is assumed to be $-121.4$ dBm \cite{yilmaz2013optimization}. Table \ref{tab:assumptions} provides a summary of these parameters as well as the offset values used in the simulations. Here it should be noted that while the hysteresis/offset values for measurement-based approaches are taken from the literature (and live network), the values for the alternative \ac{HO} methods are devised heuristically after observing the ranges of these variables in the simulations.

\par The performance of the conventional and alternative \ac{HO} triggering methods are compared in terms of the number of \acp{HO}, \acp{PPHO}, and \acp{RLF}. A \acp{PPHO} is considered to have occurred if a \ac{UE} is handed over to the target cell but returns to the original serving cell within $5$ seconds \cite{ahmad2017handover}. An \ac{RLF}, on the other hand, occurs if the \ac{SINR} of the \ac{UE} falls below a certain threshold, $Q_{out}$, and does not exceed another threshold, $Q_{in}$, (where $Q_{in} > Q_{out}$) within the time duration defined by timer T$310$ \cite{yilmaz2013optimization}. For \ac{SINR} calculation, only the interference from the other two satellites (apart from the one the user is connected to) is considered. 

\par The performance of measurement-based \acp{HO} is summarized in Table \ref{tab:measHOs} for different values of hysteresis/offset margins and \ac{TTT} parameter. As expected, the number of \acp{HO} decreases, and \acp{RLF} increase with the increase in hysteresis/offset margins and \ac{TTT} values. However, it is interesting to note that for both, static and mobile \acp{UE} the performance of the system is nearly identical. This is because the mobility of \ac{LEO} satellite itself is much higher than the user mobility ($7.56$ km/s vs $10$ m/s). Additionally, there are certain values of the \ac{HO} parameters (TTT = $20$, $Q_{hyst} + Q_{hyst} = 3$; TTT = $40$, $Q_{hyst} + Q_{hyst} = 2$; TTT = $60$, $Q_{hyst} + Q_{hyst} = 1$) where the number of both ping-pong \acp{HO} and \acp{RLF} are zero. Table \ref{tab:altHOs} evaluates the alternative \ac{HO} mechanisms described in Section \ref{sec:altHO}. Similar to the measurement-based approach, the number of \acp{HO} decreases, and \acp{RLF} increase with the increase in offset values for distance and elevation angle-based methods. In fact, for $\alpha_{off} = 10^{\circ}$ the number of \acp{HO} becomes zero. However, as mentioned earlier, this comes at the cost of increased \acp{RLF}. Additionally, it is interesting to note that even at low offset values, there is no ping-pong effect. This is due to the fact that unlike \ac{RSS}, where there is some randomness in the form of shadow fading, there is negligible randomness (due to the disparity in mobility levels of \acp{UE} and satellites) effect for distance, angle or timer-based approaches. Moreover, it is observed that for zero ping-pong and \ac{RLF} cases, the number of \acp{HO} in the alternative approaches is significantly higher compared to the measurement-based approach. This indicates that the conventional \ac{HO} triggering scheme still performs well in an \ac{NTN} scenario and can be used in the next generation \acp{NTN}.

\begin{table}[t!]
\centering
\caption{Performance comparison of measurement-based handover scheme for different hysteresis/offset margins and \ac{TTT} values}
\label{tab:measHOs}
\resizebox{\columnwidth}{!}{
\begin{tabular}{|c|c|ccc|ccc|}
\hline
                               &   & \multicolumn{3}{c|}{\textbf{Static User}}                       & \multicolumn{3}{c|}{\textbf{Mobile User}}                      \\ \cline{3-8} 
\multirow{-2}{*}{\textbf{\begin{tabular}[c]{@{}c@{}}TTT\\ (ms)\end{tabular}}} &
  \multirow{-2}{*}{\textbf{\begin{tabular}[c]{@{}c@{}} $Q_{hyst}$ + $Q_{off}$ \\ (dB)\end{tabular}}} &
  \multicolumn{1}{c|}{\textbf{HOs}} &
  \multicolumn{1}{c|}{\textbf{PP HOs}} &
  \textbf{RLFs} &
  \multicolumn{1}{c|}{\textbf{HOs}} &
  \multicolumn{1}{c|}{\textbf{PP HOs}} &
  \textbf{RLFs} \\ \hline
                               & 1 & \multicolumn{1}{c|}{28426} & \multicolumn{1}{c|}{13318} & 0    & \multicolumn{1}{c|}{28550} & \multicolumn{1}{c|}{13386} & 0    \\ \cline{2-8} 
                               & 2 & \multicolumn{1}{c|}{9966}  & \multicolumn{1}{c|}{16}    & 0    & \multicolumn{1}{c|}{9994}  & \multicolumn{1}{c|}{26}    & 0    \\ \cline{2-8} 
                               & 3 & \multicolumn{1}{c|}{9606}  & \multicolumn{1}{c|}{0}     & 0    & \multicolumn{1}{c|}{9615}  & \multicolumn{1}{c|}{0}     & 0    \\ \cline{2-8} 
\multirow{-4}{*}{\textbf{20}}  & 4 & \multicolumn{1}{c|}{9291}  & \multicolumn{1}{c|}{0}     & 5    & \multicolumn{1}{c|}{9307}  & \multicolumn{1}{c|}{0}     & 6    \\ \hline
                               & 1 & \multicolumn{1}{c|}{9719}  & \multicolumn{1}{c|}{8}     & 0    & \multicolumn{1}{c|}{9738}  & \multicolumn{1}{c|}{3}     & 0    \\ \cline{2-8} 
                               & 2 & \multicolumn{1}{c|}{9411}  & \multicolumn{1}{c|}{0}     & 0    & \multicolumn{1}{c|}{9450}  & \multicolumn{1}{c|}{0}     & 0    \\ \cline{2-8} 
                               & 3 & \multicolumn{1}{c|}{8944}  & \multicolumn{1}{c|}{0}     & 50   & \multicolumn{1}{c|}{8941}  & \multicolumn{1}{c|}{0}     & 62   \\ \cline{2-8} 
\multirow{-4}{*}{\textbf{40}}  & 4 & \multicolumn{1}{c|}{7784}  & \multicolumn{1}{c|}{0}     & 1211 & \multicolumn{1}{c|}{7749}  & \multicolumn{1}{c|}{0}     & 1253 \\ \hline
 &
  1 &
  \multicolumn{1}{c|}{9421} &
  \multicolumn{1}{c|}{0} &
  \multicolumn{1}{c|}{0} &
  \multicolumn{1}{c|}{9422} &
  \multicolumn{1}{c|}{0} &
  0 \\ \cline{2-8} 
                               & 2 & \multicolumn{1}{c|}{9009}  & \multicolumn{1}{c|}{0}     & 29   & \multicolumn{1}{c|}{8985}  & \multicolumn{1}{c|}{0}     & 30   \\ \cline{2-8} 
                               & 3 & \multicolumn{1}{c|}{7931}  & \multicolumn{1}{c|}{0}     & 1009 & \multicolumn{1}{c|}{7915}  & \multicolumn{1}{c|}{0}     & 1025 \\ \cline{2-8} 
\multirow{-4}{*}{\textbf{60}}  & 4 & \multicolumn{1}{c|}{1668}  & \multicolumn{1}{c|}{0}     & 4105 & \multicolumn{1}{c|}{1674}  & \multicolumn{1}{c|}{0}     & 4092 \\ \hline
                               & 1 & \multicolumn{1}{c|}{9188}  & \multicolumn{1}{c|}{0}     & 1    & \multicolumn{1}{c|}{9194}  & \multicolumn{1}{c|}{0}     & 2    \\ \cline{2-8} 
                               & 2 & \multicolumn{1}{c|}{8482}  & \multicolumn{1}{c|}{0}     & 307  & \multicolumn{1}{c|}{8521}  & \multicolumn{1}{c|}{0}     & 317  \\ \cline{2-8} 
                               & 3 & \multicolumn{1}{c|}{3869}  & \multicolumn{1}{c|}{0}     & 3230 & \multicolumn{1}{c|}{3966}  & \multicolumn{1}{c|}{0}     & 3228 \\ \cline{2-8} 
\multirow{-4}{*}{\textbf{80}}  & 4 & \multicolumn{1}{c|}{131}   & \multicolumn{1}{c|}{0}     & 4615 & \multicolumn{1}{c|}{137}   & \multicolumn{1}{c|}{0}     & 4630 \\ \hline
                               & 1 & \multicolumn{1}{c|}{8953}  & \multicolumn{1}{c|}{0}     & 29   & \multicolumn{1}{c|}{8919}  & \multicolumn{1}{c|}{0}     & 22   \\ \cline{2-8} 
                               & 2 & \multicolumn{1}{c|}{7517}  & \multicolumn{1}{c|}{0}     & 1241 & \multicolumn{1}{c|}{7611}  & \multicolumn{1}{c|}{0}     & 1264 \\ \cline{2-8} 
                               & 3 & \multicolumn{1}{c|}{1004}  & \multicolumn{1}{c|}{0}     & 4332 & \multicolumn{1}{c|}{997}   & \multicolumn{1}{c|}{0}     & 4336 \\ \cline{2-8} 
\multirow{-4}{*}{\textbf{100}} & 4 & \multicolumn{1}{c|}{9}     & \multicolumn{1}{c|}{0}     & 4675 & \multicolumn{1}{c|}{16}    & \multicolumn{1}{c|}{0}     & 4671 \\ \hline
\end{tabular}
}
\end{table}

\begin{table}[t!]
\centering
\caption{Performance comparison of different handover mechanisms}
\label{tab:altHOs}
\resizebox{\columnwidth}{!}{
\begin{tabular}{|c|c|ccc|ccc|}
\hline
\multirow{12}{*}{\textbf{\begin{tabular}[c]{@{}c@{}}Distance-\\Based\\ $(km)$\end{tabular}}} &
  \multirow{2}{*}{\textbf{Offset}} &
  \multicolumn{3}{c|}{\textbf{Static User}} &
  \multicolumn{3}{c|}{\textbf{Mobile User}} \\ \cline{3-8} 
 &
   &
  \multicolumn{1}{c|}{\textbf{HOs}} &
  \multicolumn{1}{c|}{\textbf{PP HOs}} &
  \textbf{RLFs} &
  \multicolumn{1}{c|}{\textbf{HOs}} &
  \multicolumn{1}{c|}{\textbf{PP HOs}} &
  \textbf{RLFs} \\ \cline{2-8} 
 & \textbf{1}            & \multicolumn{1}{c|}{14655} & \multicolumn{1}{c|}{0} & 0    & \multicolumn{1}{c|}{14656} & \multicolumn{1}{c|}{0} & 0    \\ \cline{2-8} 
 & \textbf{1.5}          & \multicolumn{1}{c|}{13524} & \multicolumn{1}{c|}{0} & 0    & \multicolumn{1}{c|}{13523} & \multicolumn{1}{c|}{0} & 0    \\ \cline{2-8} 
 & \textbf{2}            & \multicolumn{1}{c|}{11149} & \multicolumn{1}{c|}{0} & 0    & \multicolumn{1}{c|}{11149} & \multicolumn{1}{c|}{0} & 0    \\ \cline{2-8} 
 & \textbf{2.5}          & \multicolumn{1}{c|}{9572}  & \multicolumn{1}{c|}{0} & 255  & \multicolumn{1}{c|}{9572}  & \multicolumn{1}{c|}{0} & 271  \\ \cline{2-8} 
 & \textbf{3}            & \multicolumn{1}{c|}{8594}  & \multicolumn{1}{c|}{0} & 970  & \multicolumn{1}{c|}{8594}  & \multicolumn{1}{c|}{0} & 1012 \\ \cline{2-8} 
 & \textbf{3.5}          & \multicolumn{1}{c|}{7990}  & \multicolumn{1}{c|}{0} & 1627 & \multicolumn{1}{c|}{7991}  & \multicolumn{1}{c|}{0} & 1684 \\ \cline{2-8} 
 & \textbf{4}            & \multicolumn{1}{c|}{7830}  & \multicolumn{1}{c|}{0} & 2194 & \multicolumn{1}{c|}{7830}  & \multicolumn{1}{c|}{0} & 2254 \\ \cline{2-8} 
 & \textbf{4.5}          & \multicolumn{1}{c|}{7648}  & \multicolumn{1}{c|}{0} & 2547 & \multicolumn{1}{c|}{7648}  & \multicolumn{1}{c|}{0} & 2598 \\ \cline{2-8} 
 & \textbf{5}            & \multicolumn{1}{c|}{7385}  & \multicolumn{1}{c|}{0} & 2893 & \multicolumn{1}{c|}{7385}  & \multicolumn{1}{c|}{0} & 2969 \\ \hline
\multirow{10}{*}{\textbf{\begin{tabular}[c]{@{}c@{}}Elevation\\ Angle-\\ Based\\ $(\alpha)$\end{tabular}}} &
  \textbf{1$^{\circ}$} &
  \multicolumn{1}{c|}{15460} &
  \multicolumn{1}{c|}{0} &
  0 &
  \multicolumn{1}{c|}{15459} &
  \multicolumn{1}{c|}{0} &
  0 \\ \cline{2-8} 
 & \textbf{2$^{\circ}$}  & \multicolumn{1}{c|}{14990} & \multicolumn{1}{c|}{0} & 0    & \multicolumn{1}{c|}{14990} & \multicolumn{1}{c|}{0} & 0    \\ \cline{2-8} 
 & \textbf{3$^{\circ}$}  & \multicolumn{1}{c|}{14273} & \multicolumn{1}{c|}{0} & 0    & \multicolumn{1}{c|}{14274} & \multicolumn{1}{c|}{0} & 0    \\ \cline{2-8} 
 & \textbf{4$^{\circ}$}  & \multicolumn{1}{c|}{13232} & \multicolumn{1}{c|}{0} & 0    & \multicolumn{1}{c|}{13231} & \multicolumn{1}{c|}{0} & 0    \\ \cline{2-8} 
 & \textbf{5$^{\circ}$}  & \multicolumn{1}{c|}{7831}  & \multicolumn{1}{c|}{0} & 2191 & \multicolumn{1}{c|}{7831}  & \multicolumn{1}{c|}{0} & 2250 \\ \cline{2-8} 
 & \textbf{6$^{\circ}$}  & \multicolumn{1}{c|}{7501}  & \multicolumn{1}{c|}{0} & 2739 & \multicolumn{1}{c|}{7501}  & \multicolumn{1}{c|}{0} & 2808 \\ \cline{2-8} 
 & \textbf{7$^{\circ}$}  & \multicolumn{1}{c|}{6971}  & \multicolumn{1}{c|}{0} & 3316 & \multicolumn{1}{c|}{6971}  & \multicolumn{1}{c|}{0} & 3351 \\ \cline{2-8} 
 & \textbf{8$^{\circ}$}  & \multicolumn{1}{c|}{6175}  & \multicolumn{1}{c|}{0} & 3857 & \multicolumn{1}{c|}{6175}  & \multicolumn{1}{c|}{0} & 3837 \\ \cline{2-8} 
 & \textbf{9$^{\circ}$}  & \multicolumn{1}{c|}{4952}  & \multicolumn{1}{c|}{0} & 4297 & \multicolumn{1}{c|}{4952}  & \multicolumn{1}{c|}{0} & 4308 \\ \cline{2-8} 
 & \textbf{10$^{\circ}$} & \multicolumn{1}{c|}{0}     & \multicolumn{1}{c|}{0} & 4679 & \multicolumn{1}{c|}{0}     & \multicolumn{1}{c|}{0} & 4678 \\ \hline
\multirow{10}{*}{\textbf{\begin{tabular}[c]{@{}c@{}}Timer-\\Based\\ $(s)$\end{tabular}}} &
  \textbf{6.4} &
  \multicolumn{1}{c|}{15708} &
  \multicolumn{1}{c|}{0} &
  0 &
  \multicolumn{1}{c|}{15708} &
  \multicolumn{1}{c|}{0} &
  0 \\ \cline{2-8} 
 & \textbf{6.45}         & \multicolumn{1}{c|}{15708} & \multicolumn{1}{c|}{0} & 0    & \multicolumn{1}{c|}{15708} & \multicolumn{1}{c|}{0} & 0    \\ \cline{2-8} 
 & \textbf{6.5}          & \multicolumn{1}{c|}{15708} & \multicolumn{1}{c|}{0} & 0    & \multicolumn{1}{c|}{15708} & \multicolumn{1}{c|}{0} & 0    \\ \cline{2-8} 
 & \textbf{6.55}         & \multicolumn{1}{c|}{15708} & \multicolumn{1}{c|}{0} & 0    & \multicolumn{1}{c|}{15708} & \multicolumn{1}{c|}{0} & 0    \\ \cline{2-8} 
 & \textbf{6.6}          & \multicolumn{1}{c|}{15708} & \multicolumn{1}{c|}{0} & 0    & \multicolumn{1}{c|}{15708} & \multicolumn{1}{c|}{0} & 0    \\ \cline{2-8} 
 & \textbf{6.65}         & \multicolumn{1}{c|}{15706} & \multicolumn{1}{c|}{0} & 0    & \multicolumn{1}{c|}{15706} & \multicolumn{1}{c|}{0} & 0    \\ \cline{2-8} 
 & \textbf{6.7}          & \multicolumn{1}{c|}{15699} & \multicolumn{1}{c|}{0} & 0    & \multicolumn{1}{c|}{15699} & \multicolumn{1}{c|}{0} & 0    \\ \cline{2-8} 
 & \textbf{6.75}         & \multicolumn{1}{c|}{15690} & \multicolumn{1}{c|}{0} & 0    & \multicolumn{1}{c|}{15690} & \multicolumn{1}{c|}{0} & 0    \\ \cline{2-8} 
 & \textbf{6.8}          & \multicolumn{1}{c|}{15681} & \multicolumn{1}{c|}{0} & 0    & \multicolumn{1}{c|}{15681} & \multicolumn{1}{c|}{0} & 0    \\ \hline
\end{tabular}
}
\end{table}

\section{Conclusion and Future Directions}
\label{sec:Conclusion}
The increasing need for ubiquitous connectivity, coupled with  cost-effective satellite solutions driven by private ventures has led to a flurry of activity around \acp{NTN} from an industrial as well as standardization perspective. However, \acp{NTN} and particularly \ac{LEO} satellites introduce additional challenges including mobility of the satellite itself, leading to increased \acp{HO}. Furthermore, the reduced \ac{RSS} variation in these networks compared to their terrestrial counterparts calls for investigation of new \ac{HO} mechanisms. In this paper, we have compared the performance of conventional measurement-based \acp{HO} with alternatives such as location/distance, elevation angle, and timer-based methods in terms of numbers of \acp{HO}, \acp{PPHO} and \acp{RLF}. Our analysis indicates that the conventional \ac{HO} mechanism can still outperform these alternatives provided that proper selection of parameters is made.
However, here we would like to reiterate that for the purpose of this work we have only considered a homogeneous \ac{NTN}, i.e, the terrestrial network or satellites at different altitudes/orbits have not been considered. Additionally, we have not looked at the feasibility or the possible errors in obtaining the measurements such as elevation angle, location or distance. Rather, these aspects are left as a future study.

\section*{Acknowledgment}
The authors would like to acknowledge and express their gratitude for valuable suggestions provided by A. E. Duranay and M. \.{I}. Sa\u{g}lam during the preparation of this manuscript.

\bibliographystyle{IEEEtran}

\end{document}